\documentclass[twocolumn,showpacs]{revtex4}
\usepackage{graphicx}

\begin{document}

\title{The elusive source of quantum effectiveness}
\author{Vlatko Vedral}
\affiliation{Clarendon Laboratory, University of Oxford, Parks Road, Oxford OX1 3PU, United Kingdom\\Centre for Quantum Technologies, National University of Singapore, 3 Science Drive 2, Singapore 117543\\
Department of Physics, National University of Singapore, 2 Science Drive 3, Singapore 117542}

\begin{abstract}
We discuss two qualities of quantum systems: various correlations existing between their subsystems and the distingushability of different quantum states. This is then applied to analysing quantum information processing.
While quantum correlations, or entanglement, are clearly of paramount importance for efficient pure state manipulations, mixed states present a much richer arena and reveal a more subtle interplay between correlations and distinguishability. The current work explores a number of issues related with identifying the important ingredients needed for quantum information processing. We discuss the Deutsch-Jozsa algorithm, the Shor algorithm, the Grover algorithm and the power of a single qubit class of algorithms. One section is dedicated to cluster states where entanglement is crucial, but its precise role is highly counter-intuitive. Here we see that distinguishability becomes a more useful concept.
\end{abstract}

\maketitle

\section{Introduction: Classical and Quantum Correlations}

Correlations are ubiquitous in nature. The future observations we make of a system under study will in general be dependent on our past observations and the knowledge we have extracted based on them. Although we do not generally understand why events we observe around us are correlated in the first place, correlations themselves are very simply quantified within the framework of Shannon's information theory \cite{Shannon}. Suppose we perform measurements on a given system (or set of systems) repeatedly at different instants of time, $t_1, t_2,...t_N$. Let us record the outcomes of our observations as a sequence $x_1, x_2, ...x_N$. Different sequences of outcomes will naturally have different probabilities associated with them, which we will denote $p(x_1, x_2, ...x_N)$. Correlations now mean that this probability will most generally not be expressible as product of probabilities of subsequences, $p(x_1,...x_n)\times p(x_{n+1},...x_N)$ for any $1\leq n \leq N$. Shannon introduced the notion of mutual information in order to quantify how correlated different observations are. For simplicity, if we divide measurements into two groups, $A$ and $B$, each of them having a well defined probability distribution, $p(A)$ and $p(B)$ respectively, as well as a joint probability distribution, $p(A,B)$, then the mutual information between $A$ and $B$ is defined as $I(A:B) = H(A)+H(B)-H(A,B)$. Here $H(X)=-\sum_{x\in X} p(x)\log p(x)$ is the well-known Shannon entropy. There is a certain degree of subtlety in trying to extend Shannon's mutual information to more than two different sets of outcomes, but this issue will not concern us in the current exposition.

The concept of mutual information is so general that it can easily be extended to quantum systems \cite{Vedral_RMP}. This leads us to the notion of quantum mutual information, which, for a general state $\sigma_{AB}$, is defined as $I(\sigma_{AB}) = S(\sigma_A)+S(\sigma_B)-S(\sigma_{A,B})$, where $S(\rho)=-tr \rho\log \rho$ is the von Neumann entropy and $\sigma_{A}$ and $\sigma_{B}$ are the reduced density matrices of state $\sigma_{AB}$. However, in quantum mechanics, we have learnt to discriminate between different forms of correlations, a distinction that has no counterpart in classical information theory. First of all there is entanglement. Given a bipartite quantum state $\sigma_{AB}$, entanglement presents any form of correlation that cannot be captured by the states of the form $\sum_i p_i \rho_A^i\otimes \rho_B^i$ (which are known as separable or disentangled). Entanglement in $\sigma_{AB}$ is then most easily quantified by calculating how different this state is to any separable state \cite{Vedral1,Vedral2}. This difference can be expressed in a number of ways, but the related details will not trouble us at present (see, for instance, \cite{Amico}). Among the separable states, however, there are those that we can call classically correlated. This will simply mean that we have orthogonal states for subsystem $A$, call them $|k\rangle$, and orthogonal states for system $B$, $|l\rangle$, and the probabilities corresponding to them, $p_{kl}$ will not simply just be equal to $p_k\otimes p_l$. Classically correlated states would therefore have a general form $\sum_{kl} p_{kl} |k\rangle\langle k|\otimes |l\rangle\langle l|$. There are clearly separable states that are not just classically correlated in this very sense. One example is the state $\rho_{AB} = 1/2 (|0\rangle\langle 0|_A\otimes |0\rangle\langle 0|_B + |1\rangle\langle 1|_A\otimes |+\rangle\langle +|_B$, where $|+\rangle = (|0\rangle + |1\rangle)/\sqrt{2}$. It will become transparent later why we need to discriminate between separable states and classically correlated states. We can also use some entropic measure to quantify how different separable states are from classically correlated ones and this too will be discussed shortly. The states containing no correlations, either quantum or classical, are called product states, $\rho_A\otimes\rho_B$.

An equivalent way to Shannon's of quantifying correlations is to think of the reduction in entropy of $A(B)$ when $B(A)$ is measured. The more correlated $A$ and $B$, the more we can learn about one of them by measuring the other.
Suppose we make measurements on $A$. For each measurement outcome $i$, occurring with probability $p_i$, the state of $B$ will collapse to $\rho_B^i$. Classical correlations in a state $\rho_{AB}$ are then simply the maximum over all measurements performed on $A$ of the quantity $C(\rho_{AB}) = S(\rho_B) - \sum_i p_i S(\rho_B^i)$ (as defined in \cite{Henderson}; see also \cite{Vedral-PRL}). We can also define this quantity by swapping the roles of $A$ and $B$, but the subtleties related to the question of symmetry of classical correlations will be not relevant for our present discussion.

For classically correlated states it is clear that their quantum mutual information and classical correlations lead to exactly the same measure of correlations. What is rather intriguing, however, is that for separable states the mutual information is generally larger than classical correlations. This means that separable states contain correlations over and above just the classical ones. The discrepancy between the two is known as the quantum discord, $D=I-C$. We will call discord the correlations over and above classical, but excluding entanglement. (Note: in \cite{Zurek} the discord is defined to contain entanglement as well).

The general picture is this. Quantum mutual information in any quantum state $\sigma_{AB}$ can be written as $I=E+C+D$, where $E$ is the amount of entanglement in the state (as measured by the relative entropy of entanglement \cite{Vedral1} to make it on an equal footing with other entropic measures of correlations). Physically this means that the quantum mutual information measures total correlation in a quantum state, which can be though of as consisting of entanglement, $E$, classical correlations, $C$, and the additional quantum correlations, $D$, which are not due to entanglement. For pure states, discord always vanishes and total correlations are conveniently equal to the sum of entanglement and classical correlations \cite{Henderson}. Moreover, both entanglement and classical correlations in this case are equal to one another and to either of the reduced von Neumann entropies. This is an expected consequence of the Schmidt decomposition of pure bipartite states \cite{Vedral_RMP}. For mixed states, discord is generally non-vanishing, and this seems to hold important implications for quantum information processing, the topic of main focus in the present paper.

\section{Information Processing and Distinguishability}

To motivate the forthcoming discussion we first ask the question: what feature of quantum mechanics makes quantum information processing more efficient than classical? It has frequently been said that entanglement is clearly that feature. The answer seems obvious in the case of pure states. If there is no entanglement (or very little of it) during the evolution of pure states, then that evolution can efficiently (with only a polynomial overhead) be simulated by classical systems \cite{Jozsa}. But, we should remember that according to our above discussion, pure states contain the same amount of classical correlations as entanglement. Therefore, we might well say that it is classical correlations in pure states that are responsible for the speed-up! The picture, however, changes dramatically for mixed states. First of all, any evolution of just classically correlated states can be simulated by classical computers (by definition, for these state define what we mean by classical computers); therefore classical correlations cannot be, on their own, responsible for the speed-up. Furthermore, it is possible to have a speed-up with just separable (more than classically correlated) states and this means that entanglement cannot be responsible for the speed-up either. If we look among correlations for the culprit, then we are only left with the discord, which is non-zero for mixed separable, but non-classically correlated states. This conclusion, however, would not be consistent with the pure state analysis, where discord is non-existent. We are finally left in an uncomfortable position: none of the correlations, quantum or classical, can (singularly) be responsible for the speed-up of quantum information processing!

Besides correlations, we have the concept of indistinguishability in quantum mechanics, namely the fact that different quantum states, unlike classical, need not, even in principle, be distinguishable from one another. This fact is key in quantum communications in general, and quantum cryptography in particular. The fact that Alice encodes two messages, $0$ and $1$ into non-orthogonal quantum states $|0\rangle$ and $|+\rangle$, makes it impossible for any eavesdropper to remain un-noticed. Information cannot be extracted from non-orthogonal states without disturbing them. In fact, quantum computation can also be viewed as a form of information processing where different, in general non-orthogonal, outputs have to be discriminated from each other (see \cite{Bose} for an exposition of this view). This is why we might expect that separable states with non-zero discord could still be more efficient than just classically correlated states.

To illustrate how discrimination enters computation, let us look at the concrete example of the Deutsch-Jozsa algorithm \cite{Deutsch}. This particular problem achieves an exponential speed-up over classical problems. Here, we are promised a function which is either constant (all outputs are $0$ or all are $1$) or balanced (outputs contain an equal number of zeros and ones). If we are restricted to a single application of the function, it is clear that classically we cannot obtain any information. Knowing the value of a single bit out of $N$ bits, implies no knowledge of the rest $N-1$ bits. Quantumly, however, we can think of the evaluation of the function on $x$ as the implementation of the phase  factor $e^{i\pi f(x)}$. Then, if we input the state $|+\rangle\otimes|+\rangle\otimes...|+\rangle$, where $|+\rangle = |0\rangle +|1\rangle$, the output will either be the $\pm |+\rangle\otimes|+\rangle\otimes...|+\rangle$ if the function is constant, or it will be one of the orthogonal states containing the superposition of all states with half of the phase factors negative. Thus the final measurement is a simple orthogonal, projective measurement to discriminate the two case. It can be shown that here entanglement will exist in general among the states resulting from the application of a balanced function\cite{Azuma}. We will therefore turn to mixed states in order to show that entanglement is not needed for the higher quantum efficiency.

Imagine that the input state is a mixture of the maximally mixed state on $N$ qubits, with the corresponding probability $1-\epsilon$, and the pure state $|+\rangle\otimes|+\rangle\otimes...|+\rangle$, with the probability $\epsilon$. These states arise, for example, in the liquid state Nuclear Magnetic Resonance quantum information processing, and we will henceforth refer to them as pseudopure (I am being a bit cavalier with mathematics here: the natural states in NMR are the thermal Gibbs state, but they could, for all practical purposes, be approximated well by pseudopure states. Pseudopure state are mathematically easier to handle which is their chief appeal). Providing that $\epsilon <1/(2^{2n-1}+1)$, this state will never become entangled under any unitary evolution \cite{Braunstein} (and hence any functional evaluation in the Deutsch-Jozsa algorithm, for example) since it is sufficiently mixed that a separable decomposition is always possible. However, no matter how small the $\epsilon$, we can show that some non-zero information can still be obtained regarding the nature of the function. This is because the resulting (output) two mixed states, corresponding to the constant and the balanced function respectively, can always be partly discriminated. How much information can be obtained is conveniently quantified by the Holevo bound. This looks at the difference between the entropy of the mixture of the two states minus the average of the entropies of the individual states.

Suppose that the probability with which we are given a balanced function is $p$ (and so the probability for the constant function is $1-p$). Then the final state of the computer is
\begin{eqnarray}
\rho_f & = & p U_b(1-\epsilon)I+\epsilon |+\rangle\langle +|^{\otimes n})U_b^{\dagger} \\
& + & (1-p)U_c(1-\epsilon)I+\epsilon |+\rangle\langle +|^{\otimes n})U_c^{\dagger}
\end{eqnarray}
where $U_{b,c}$ are the unitary transformations implementing the balanced and the constant function respectively.
The information we can now extract about the nature of the function is
\begin{eqnarray}
I^{out} & = & S(\rho_f) - p S(U_b(1-\epsilon)I+\epsilon |+\rangle\langle +|^{\otimes n})U_b^{\dagger}) \\
& - & (1-p)S(U_c(1-\epsilon)I+\epsilon |+\rangle\langle +|^{\otimes n})U_c^{\dagger})
\end{eqnarray}
We can show that in the limit of small $\epsilon$, which is what we require to make the state always separable, this quantity scales as $I^{out} \approx 2^n \epsilon^2 +O(\epsilon^3)$ \cite{Mor}.

Let us compare the output information gain to the correlations of one qubit with the rest $N-1$ qubits. Classically we can only measure one bit, and we said that this would give us no information about the nature of the function. This is because the state of one bit is in no way correlated to whether we have applied $U_b$ or $U_c$. Quantumly, however, although there is no entanglement in the final state, we do have a finite discord. Again in the limit of small $\epsilon$ the discord is calculated to be $D \approx 2^n \epsilon^2 +O(\epsilon^3)$ (to be shown in the next section), and is therefore directly related to the information we obtain, $I^{out}$.

It is important to stress that the relationship between information out and the discord only holds under the assumption that $\epsilon$ is small (so that we can make a Taylor expansion of various entropies to their lowest order). We will see in the next section that this translates into $\epsilon <<2^{-n}$. This limit is in perfect accord with the fact that we require $\epsilon < 1/(2^{2n-1}+1)$ in order to guarantee separable states. (At the other extreme, when we consider pure states, we have already noted that the speed-up occurs even tough the discord is always zero).

The relationship between information obtained and discord for highly mixed states is not accidental. We now proceed to show its exact form for general promise type problems.

\section{Distinguishability and Discord}

Suppose we start with a pseudopure mixed state of $n$ qubits, $\rho_n = (1-\epsilon)I/2^n + \epsilon |0\rangle\langle 0|^{\otimes n}$. Furthermore, imagine that we are promised $N$ different properties encoded into unitaries $U_1,U_2...U_N$, with the respective probabilities $p_1,p_2,...p_N$. The amount of information we can obtain at the end about different properties is, as we have seen, given by $I^{out} = S(\sum_i p_i \rho_i) - \sum_i p_i S(\rho_i)$ where $\rho_i = U_i\rho_n U_i^{\dagger}$. This expression can be simplified since $S(\rho_i)$ has the same value for all the outcomes (because they are all unitarily related with the input state and hence must have the same entropy as the input state).

Let us now look at the discord between the first qubit and the rest in the final state after one unitary has been applied. It is given by
\begin{eqnarray}
D= 1 - S(\rho_n) + S(\rho_{n-1}) \; .
\end{eqnarray}
We can easily compute both $S(\rho_n)$ and $S(\rho_{n-1})$,
\begin{eqnarray}
S(\rho_n) & = & -(\frac{1-\epsilon}{2^n} + \epsilon)\log (\frac{1-\epsilon}{2^n} + \epsilon)\\
& - & (2^n-1)\times \frac{1-\epsilon}{2^n}\log (\frac{1-\epsilon}{2^n}) \\
&\approx & n - 2^n \epsilon^2 \; .
\end{eqnarray}
By the same token $S(\rho_n)\approx n-1 - 2^{n-1} \epsilon^2$. The discord is now, to the lowest order in
$\epsilon$ equal to
\begin{eqnarray}
D \approx 2^{n-1} \epsilon^2 \; .
\end{eqnarray}
The mutual information, on the other hand, is calculated to be
\begin{eqnarray}
I^{out} & = & S(\sum_i p_i \rho_i) - S(\rho_i) \leq n - S(\rho_i)  \\
& \approx & n - (n-2^n\epsilon^2) = 2^n \epsilon^2 = 2D
\end{eqnarray}
Therefore, here we have a general inequality that the amount of information we can extract, which tells us about the efficiency of our quantum information processing, is bounded by (twice) the discord. This immediately shows that if the discord is zero, then no information can be obtained within this framework.

In computing the discord we have assumed that the final pure state in the mixture is of the form $|0\rangle\otimes |\Psi_0\rangle + |1\rangle\otimes |\Psi_1\rangle$. The states $|\Psi_0\rangle$ and $|\Psi_1\rangle$ need not be orthogonal in general, though in the Deutsch-Jozsa algorithm they certainly are in the case of the balanced function (this is true when three and more qubits are concerned; for two qubits there is no entanglement present anywhere, at any stage of computation).

Note, again, that this is in no contradiction with the fact that pure states have zero discord and yet lead to a quantum advantage. The reason is that we were assuming here that $\epsilon$ is so small (because we insisted on separability of all states involved) that we could make an approximation to both the discord as well as the information gain. For nearly pure states this approximation, of course, fails, and the discord is no longer an appropriate upper bound.

Can anything be said about more general algorithms? We have so far only required a better than classical efficiency.
What happens if we are a bit more demanding and ask for a significant difference in efficiency?

\section{Discussion}

We can now raise the bar and ask for quantum protocols that are not just more efficient than classical, but exponentially so (though at the end of the section we will see that even a polynomial speedup can in some cases be addressed by similar means). Exponential efficiency means that the time it takes to reach the answer scales quantumly as a polynomial of the number of (qu)bits required for memory, while it takes an exponential time for any classical computer (Note: it is important that we keep the memory polynomially bounded). We ask if entanglement is needed in this case.

We can answer this question in the affirmative in one special case \cite{Linden}. Suppose that our initial state is a pseudopure state. If the pure fraction is $\epsilon$, then the number of times we need to repeat the computation to obtain a correct result is of the order of $1/\epsilon$. If we assume, like in Deutsch-Josza, that we are computing some (non-constant) function $f$, then the pure part of the pseudopure mixture will generally evolve to be $\sum_x |x\rangle \otimes |f(x)\rangle$ (this is also true for Grover's \cite{Grover} and Shor's \cite{Shor} algorithms). Let us now look at the entanglement between the first and the second register in the pseudopure mixture. We project the first register onto the $|0\rangle,|1\rangle$ subspace (without destroying the coherence between the two states). Then, we obtain an effective two qubit pseudopure state
\begin{equation}
\rho_{2\times 2} = (1-\delta) I + \delta \frac{(|0f(0)\rangle + |1f(1)\rangle)}{\sqrt{2}}\frac{(\langle 0f(0)| + \langle 1f(1)|)}{\sqrt{2}}
\label{exponential}
\end{equation}
where $\delta = 1/((1-\epsilon)2^{-n_2+1} +\epsilon)$ and $n_2$ is the number of qubits in the second register (roughly equal to $n/2$ in general). Since this is a two qubit state, it is entangled if and only if $\delta \geq 1/3$ which implies that $\epsilon \leq 1/(2^{n_2}+1)\approx 1/2^{n/2}$. If $\epsilon$ is not in this domain than the original $n$ qubit pseudopure state must have been entangled (since we performed a local projective measurement and this, by definition, cannot create entanglement out of a separable state). To avoid entanglement, therefore, the pure fraction must not be greater than $\epsilon \approx 2^{-n/2}$ and this means that the resulting computation cannot be exponentially more efficient than classical.

The whole above discussion can conveniently be phrased in terms of distinguishability. If your input state is too distinguishable from the pure state that would yield the maximum quantum efficiency, then there is no gain in using it for quantum computing. In this case we can use the relative entropy to quantify this distinguishability \cite{Vedral3}. This is an asymptotic measure that is only achieved in the limit of large number of trials, but it otherwise provides an upper bound for any finite case scenario. Supposing again that we work with a pseudopure state, this leads to:
\begin{eqnarray}
S(|\Psi\rangle\langle \Psi| \; ||\; \epsilon |\Psi\rangle\langle \Psi| + (1-\epsilon) \frac{I}{2^n}) & = & -\log (\epsilon + \frac{1-\epsilon}{2^n}) \nonumber \\
& \approx & \log \frac{1}{\epsilon} \; .
\end{eqnarray}
The probability that the pseudopure state will be confused with $|\Psi\rangle\langle \Psi|$, which means that we have a successful outcome, is simply given by the exponential of the above relative entropy \cite{Vedral3}
\[\exp\{-S(|\Psi\rangle\langle \Psi|\; ||\; \epsilon |\Psi\rangle\langle \Psi| + (1-\epsilon) \frac{I}{2^n})\} = \epsilon \; .
\]
This is the same conclusion as above, namely we need to repeat the computation roughly $1/\epsilon$ times to have a unit success. If, furthermore, we require the pseudopure state to be distinguishable from a separable state, then $\epsilon <1/2^n$ and hence there is no exponential speed up in the resulting computation.

This argument is very simple, but what does it mean? Does it mean that, for example, Shor's algorithm definitely uses entanglement? If our input is pseudopure, then the answer is yes. However, if we use a mixture of another type the answer remains unknown (although there is numerical evidence that entanglement always appears \cite{Parker}). What is more, we have evidence that different mixtures can achieve exponential speed-up without entanglement (or with very little of it) in another instance. This instance is, in fact, also another clear example of the connection between discord and distinguishability and lies at the heart of the efficiency of quantum computation. The task is to compute a trace of a unitary matrix and can be accomplished with one non-maximally mixed qubit and a completely mixed register of $n$ qubits \cite{Knill}. Here the input state is of the form $|+\rangle\langle +|\otimes I/2^n$ (note: this is not a pseudopure state).
The said unitary is applied only if the first qubit is in the state $|1\rangle \langle 1|$. This leads to the state
\begin{equation}
\rho_U = (|0\rangle \langle 0|+|1\rangle \langle 1|)\otimes \frac{I}{2^n} + |0\rangle \langle 1|\otimes U^{\dagger} + |1\rangle \langle 0|\otimes U
\end{equation}
(The trace of $U$ is now extracted by measuring the first qubit in the $|\pm\rangle$ basis). It is clear that here there is no entanglement between the first qubit and the mixed register. Furthermore, we can make the first qubit
arbitrarily close to the mixed state (where mixedness is constant, i.e. independent of the number of qubits) and still have an exponential quantum advantage \cite{Datta}.  It is tempting therefore to look for reasons other than entanglement to explain the speedup. Discord is certainly one such measure (as proposed in \cite{Datta}), but this, again, is related to the distinguishability between the states of the mixed register resulting from measuring the qubit in the $|\pm\rangle$ basis. Therefore, here a similar link exists between speedup, discord and distinguishability that we also found in the Deutsch-Jozsa algorithm. Note that (in line with what has been said) there is evidence that this algorithm is difficult to simulate classically, even tough the overall entanglement scales poorly (or is non-existent) \cite{Vidal}.

Grover's algorithm \cite{Grover} is interesting to mention here simply because the speedup is only polynomial, i.e. quadratic, but the algorithm is of a very general nature (since any difficult problem boils down to a search). It has been established that pure state Grover's algorithm in general contains entanglement between qubits \cite{Azuma}. There have been claims that search can be done without entanglement, but this is only true, so far as we can tell, when the memory encoding is inefficient. For example, a ``classical laser beam" (meaning: a coherent state with a large average number of photons) can perform Grover's search using a diffraction grating encoding a database to be searched \cite{Fourier}. Each slit in the grating here represents a different database element. It is clear that this is less efficient than using (qu)bits to encode the database since we only need $\log N$ qubits to encode $N$ database elements whereas we need $N$ slits in the diffraction grating (i.e. exponentially more spatial resources) to do the same. Here entanglement is thus linked with efficient spatial encoding.

What happens if we use the pseudopure states to run Grover's algorithm? It has been shown that if $\epsilon \geq 1/\log N$, then Grover's search algorithm is still more efficient than any classical algorithm \cite{Biham} (this immediately follows from the fact that the probability of success is reduced by $\epsilon$ when we have pseudopure states. If $\epsilon$ scales as a polynomial of the number of qubits $n=\log N$, then we only need polynomially many repetitions to achieve Grover's quadratic speedup. The overall quantum time is therefore $\approx \sqrt{N}/\log N\approx \sqrt{N}$). However, as we have seen, these states are, in fact, entangled. The reason is that all states of the form in eq. (\ref{exponential}) also occur during Grover's algorithm (in Grover's search $f(0)=0$, and this represents all irrelevant database elements, while  $f(1)=1$ corresponds to the database element we are looking for; at some stage of the algorithm the two will have comparable amplitudes, which is all we need in this argument). Here, therefore, as soon as we require anything faster than classical (which scales as $N$) we immediately have entangled pseudopure states. We cannot, of course, rule out that there are some other mixed states that are separable and yet achieve a quadratic speedup. However, for states where some qubits are pure and others are maximally mixed (as in the case of the power of a single qubit), evidence points to the necessity for entanglement \cite{Biham}.

A general point deserves special attention in our discussion. It appears that a strong criterion for quantum effectiveness is the fact that the state throughout the computation is sufficiently different to classically correlated state (though this need not mean that it is entangled at any stage). We have seen that the relative entropy conveniently tells us about how distinguishable two states are. Suppose, therefore, that we ask a less demanding, but related, question: how distinguishable is a maximally noisy state from a given pseudopure state? This can be computed to be:
\begin{eqnarray}
& & S(\epsilon |\Psi\rangle\langle \Psi| + (1-\epsilon) \frac{I}{2^n}\;||\;\rho_c) = \\
& &   -S(\epsilon |\Psi\rangle\langle \Psi| + (1-\epsilon) \frac{I}{2^n}) - \\
& & tr \{\epsilon |\Psi\rangle\langle \Psi| + (1-\epsilon) \frac{I}{2^n} \log \rho_c\} = \\
& &  2^n \epsilon^2 + O(\epsilon^3)
\end{eqnarray}
(again assuming a small $\epsilon$ expansion). Let us now compute this same quantity, but for the power of a single qubit state and see if and by how much more this state is distinguishable from a maximally noisy one. The relative entropy is now given by:
\begin{equation}
S(U\{|0\rangle\langle 0|\otimes \frac{I}{2^n}\} U^{\dagger} \;||\;\rho_c)  = 1 \; .
\end{equation}
It is clear that the state single pure qubit state is much more distinguishable (exponentially more so!) from pure noise than the pseudopure state (bearing in mind that $\epsilon \approx 2^{-n}$). This direction will require much further research, but can we say that this kind of distinguishability is at the root of the efficiency of some states and evolutions as opposed to others? The mathematical intricacy will lie in, firstly finding the best classically correlated state to approximate our quantum state (something that I conveniently avoided doing above) and, secondly, doing this for each instance in time as the state evolves in a unitary fashion.

\section{Digression: Cluster States}

There exists a computational model exempt from above considerations in that entanglement is definitely a necessary resource for it. Here I have in mind the so called cluster state quantum computation (or measurement based quantum computation) \cite{Briegel}. This form of computation consists, in fact, of a sequence of one qubit measurements followed by a feed-forward of the information contained in the measurement outcomes. Measurements are performed on highly entanglement initial states of many qubits, the so called cluster states. Typically, parts of clusters are measured so that (as a consequence) the remaining, unmeasured, qubits undergo a desired computation. The feed-forward of classical information ensures that the evolution of the remaining part is unitary (deterministic) in spite of being driven by measurements.

Cluster state computation is, simply speaking, a generalization of the teleportation protocol to other, more complicated, algorithms. Just like in teleportation, entanglement is crucial for cluster state computation. A cluster state that is only separable cannot achieve any advantage over classical computers. However, and this is at first sight a surprising fact, too much entanglement in clusters can also be detrimental to quantum computation \cite{Eisert}. This is because a highly entangled state, when measured, tends to give the output that is indistinguishable from a maximally mixed state (note: here the distingishability is between the output bit strings made up of measurement outcomes and not between the states themselves, though the two are, of course, somewhat related). We now proceed to explain what it means to be highly entangled.

We can phrase the success probability for computation in terms of relative entropy in the following way (for a more detailed and rigorous analysis see \cite{Eisert}):
\begin{enumerate}
\item Consider the sequence of bits $b=(b_1,b_2,...b_N)$ generated by making one qubit measurements on the entangled state used for measurement based computing.
\item The probability of getting a particular string is $p(b) = |\langle b_1,b_2,...b_N|\Psi\rangle|^2$, where $|\Psi\rangle$ is the entangled state itself.
\item Let the number of bit strings giving us a non-zero probability of success be
$N_s$.
\item Then the total probability of success can be estimated to be \[p_s = \sum_{i\in N_s} |\langle b_1,b_2,...b_N|\Psi\rangle|^2\leq \sum_{i\in N_s}e^{-E_{\Psi}} = N_s 2^{-E_{\Psi}} \] where $E_{\Psi}$ is (twice the log of) the geometric measure of entanglement \cite {Wei} (this itself is a lower bound on the relative entropy of entanglement, but for cluster states the two coincide).
\end{enumerate}
Computation should only proceed if the probability of success is finite, $p_s=c>0$, so that the correct result can be achieved by repetition ($c$ is a constant independent of the number of qubits $N$). This implies that
\[
N_s\geq 2^{E_{\Psi}-\log (1/c)} \; .
\]
Now, if entanglement scales as the number of qubits $N$ (for large $N$, to be precise, entanglement in most states scales as $N-\log N$, but this logarithmic correction is immaterial in the thermodynamical limit) then it follows that the number of successful solutions to our problem is equal to the total number of outputs (minus a constant factor $\log (1/c)$ which, again, is irrelevant in the large $N$ limit). Therefore, the computation using such entangled state can be simulated using a completely random coin toss. It turns out that cluster states have exactly $N/2$ units of entanglement which clearly does not lead to a trivial result from the above inequality (any reader interested in entanglement scaling in various many-body states could consult the elementary review in \cite{Vedral-Nature}). Anything much smaller than this (e.g. $\log N$) would be insufficient as a universal resource, but for a different reason: this state, and measurements made on it, could be simulated by classical means.

It is interesting to note that entanglement in cluster states behaves very much like the free energy or entropy (hence too much entanglement in a state is akin to too high an entropy of the resulting computation). Suppose that we have a mixed state due to the system being at a finite temperature, $T$. The probability that we are in the ground state is then $p_G = e^{(-E_G+F)/kT}$ where $E_G$ is the ground state energy and $F=-kT\log Z$ is the free free energy ($Z$ being the partition function). Let us, for simplicity, assume that $E_G=0$ and that $kT=1$. The estimate for the number of successful bit strings then becomes,
\[
N_s\geq 2^{E_{\Psi}-F-\log (1/c)}
\]
(since $p_s \leq N_s p_G 2^{-E_{\Psi}}= N_s2^{-E_{\Psi}+F}$).
This shows that if $E_{\Psi}-F \approx N$, the state is useless for quantum computation. There is here obviously a tradeoff between free energy (or entropy) and entanglement. Too much entanglement simply implies too low a free energy (or too high an entropy) which has the ``effect" to generate too much noise in the output. This is (thermodynamically speaking) why the output then becomes indistinguishable from a random state. In some sense, performing cluster state quantum computation, is analogous to doing useful work, which is only possible if the state has non-zero free energy, i.e. if it is sufficiently different to a maximally mixed state (for a more in-depth discussion of analogies between clusters and thermodynamics see \cite{Anders}).

The bottom line, ultimately, is this. The fact that entanglement is needed for cluster states, does not mean that we cannot achieve significant speedups without it, simply because cluster states are just one way of executing quantum computation. Though clusters are a universal resource, entanglement in them is really a substitute for the missing unitary dynamics (given that only single qubit measurements are at our disposal). So, even here, it is not clear which resource is responsible for the quantum effectiveness. The focus would have to shift to measurements and the effect of noise on them as well as the ensuing computation.

\section{Conclusion and Outlook}

In addition to the search for the source of power of quantum information processing, there is a related physical issue about the relationship between the concepts of superposition and entanglement. It is, in fact, very difficult to discuss the cause (or causes) for the quantum information speedup, without immediately running into some fundamental physical issues (information, after all, is physical). I will use a simple example related to above discussion to illustrate the point.

Suppose that the pure qubit in the example of  ``the trace computation" (the power of a single qubit) is a photon entering an interferometer. Then, after the first beamsplitter, the photon is in the state of a superposition of two paths. However, this state can also be considered an entangled state as it is written as $|01\rangle + |10\rangle$. In fact, we have shown elsewhere that this state can violate Bell's inequalities \cite{Dunningham} and is, therefore, a legitimate (though single particle) entangled state on a par with the state $|HH\rangle+|VV\rangle$ of two photons in, say, parametric down conversion ($H,V$ stand for horizontal and vertical polarisation respectively). The subsequent unitary operation conditional on the second mode being $|1\rangle$ is then just a local unitary transformation applied to the second mode and it must thus preserve the original entanglement. In fact, even if we start with a mixed state (as long as it is not completely mixed), our resulting state will still always be entangled between the two spatial modes (since it is just given by the mixture $p|\Psi^+\rangle \langle \Psi^+| + (1-p)|\Psi^-\rangle \langle \Psi^-|$). This state has the relative entropy of entanglement equal to $E= 1 + p\log p +(1-p)\log (1-p)$ \cite{Vedral2}. Within such single pure photon implementation, entanglement is clearly always present and could therefore be said to be responsible for the speedup (as much as any classical or total correlations are).

We are at the end of our search for the source of quantum effectiveness and one conclusion can safely be drawn: we should give up looking for a single reason behind the quantum speedup. Most likely, the answer will intimately be connected with the exact nature of the problem and, as seen above, will vary from problem to problem. Though possibly intellectually displeasing, this answer is the only possible consistent one at present. This leads us to the following final thought.

Let us at the end of our investigation take a broader view of information processing. Beyond man-made computational devices, there are, of course, much older and more ubiquitous information processors in nature - the living systems. All living systems are very opportunistic (possibly even more so than theoretical physicists) and what matters to their survival is to be able to gain even the smallest available advantage over their competitors. In natural information processing a fraction of a second speedup over and above one's predator, for instance, makes all the difference in the world. Nature could not care less about exponential improvements - it simply does not see beyond the next step. Any improvement that results in a higher chance of survival will simply suffice (though in the long run, all the incremental steps may ultimately add up to an exponential improvement). Therefore, if we generalise our question and ask whether quantum physics could improve certain natural operations, it is to be expected that entanglement may no longer be the most important resource. All tricks of the quantum trade will then be exploited, very much in the spirit of the present discussion.

\textit{Acknowledgments}: This paper was stimulated by an invited talk at the QuPa meeting in Paris, delivered on $28$th May $2009$. The author acknowledges financial support from the Engineering and Physical Sciences Research Council, the Royal Society and the Wolfson Trust in UK as well as the National Research Foundation and Ministry of Education, in Singapore. The author is a fellow of Wolfson College Oxford.

\end{document}